# Using Friends as Sensors to Detect Global-Scale Contagious Outbreaks

Manuel Garcia-Herranz[1], Esteban Moro Egido[2,3,4], Manuel Cebrian[5,6,7], Nicholas A. Christakis[8,9], James H. Fowler [10,11]

[1] *Escuela Politécnica Superior, Universidad Autónoma de Madrid, Madrid 28049, Spain*
[2] *Departamento de Matematicas & GISC, Universidad Carlos III de Madrid, 28911 Leganés, Spain*
[3] *Instituto de Ingeniería del Conocimiento, Universidad Autónoma de Madrid, Madrid 28049, Spain*
[4] *Instituto de Ciencias Matemáticas CSIC-UAM-UCM-UC3M (ICMAT), 28049, Madrid, Spain*
[5] *Computer Science & Engineering Department, University of California, San Diego, CA 92093, USA*
[6] *Media Laboratory, Massachusetts Institute of Technology, Cambridge, MA 02139, USA*
[7] *National Information and Communications Technology Australia, Melbourne, Victoria 3010, Australia*
[8] *Department of Sociology, Harvard University, Cambridge, MA 02138, USA*
[9] *Department of Medicine, Harvard Medical School, Boston, MA 02115, USA*
[10] *Medical Genetics Division, School of Medicine, University of California, San Diego, CA 92103, USA*
[11] *Political Science Department, University of California, San Diego, CA 92103, USA*

*Corresponding Author: JHF, jhfowler@ucsd.edu

**Recent research has focused on the monitoring of global-scale online data for improved detection of epidemics[1,2,3,4], mood patterns[5,6], movements in the stock market[7] political revolutions[8], box-office revenues[9], consumer behaviour[3,10] and many other important phenomena. However, privacy considerations and the sheer scale of data available online are quickly making global monitoring infeasible, and existing methods do not take full advantage of local network structure to identify key nodes for monitoring. Here, we develop a model of the contagious spread of information in a global-scale, publicly-articulated social network and show that a simple method can yield not just early detection, but advance warning of contagious outbreaks. In this method, we randomly choose a small fraction of nodes in the network and then we randomly choose a "friend" of each node to include in a group for local monitoring. Using six months of data from most of the full Twittersphere, we show that this friend group is more central in the network and it helps us to detect viral outbreaks of the use of novel hashtags about 7 days earlier than we could with an equal-sized randomly chosen group. Moreover, the method actually works better than expected due to network structure alone because highly central actors are both more active and exhibit increased diversity in the information they transmit to others. These results suggest that local monitoring is not just more efficient, it is more effective, and it is possible that other contagious processes in global-scale networks may be similarly monitored.**

Modern social, informational, and transactional platforms offer a means for information to spread naturally (e.g, as in the case of the "Arab Spring"[11]), and there is increasing interest in using these systems to intentionally promote the spread of information and behaviour[12,13,14,15]. They they also yield a brand-new and large-scale global view of social interactions and dynamics of formerly hidden phenomena[16]. However, the advent of global monitoring has recently heightened concerns about privacy[17] and scholars have shown that anonymization is insufficient to guarantee it[18]. Thus, future efforts to monitor global phenomena may be restricted to analysis at a local scale[4,19]. Moreover, the explosive growth of online data has made it more and more difficult to perform a complete global analysis. As a result, scholars are beginning to develop local methods that sample small but relevant parts of the system[20,21].

Here, we elaborate a new sampling technique that takes advantage of the local structure inherent in large-scale online social networks, and we use it to test an important hypothesis about social contagion. If a message is transmitted via *broadcast*, then all individuals in a network are equally likely to receive it. On the other hand, if a message is transmitted from person to person via *contagion*, then individuals at the centre of a network are likely to receive it sooner than randomly-chosen members of the population because central individuals are a smaller number of steps (degrees of separation) away from the average individual in the network[22,23]. As a result, for contagious processes, we would expect the S-shaped cumulative "epidemic curve"[24] to be shifted to the left (forward in time) for centrally located individuals compared to the population as a whole.

If so, then the careful collection of information from a sample of central individuals within human social networks could be used to detect contagious outbreaks before they happen in the population at large[22]. We call this the *sensor hypothesis*. In fact, the very discrepancy in the time to infection between central and randomly-chosen individuals could serve as a means to distinguish between broadcast and contagion mechanisms, either *ex post* by comparing their mean times of infection or in real time by looking for the first day in which there is a significant difference in their cumulative incidences.

Using 6 months of data from Twitter recorded in 2009, we analyse a network containing 40 million users around the world who are connected by 1.5 billion directed relationships ("follows"). Over six months, these users sent nearly half a billion messages ("tweets"), of which 67 million contained a user-supplied topic keyword called a "hashtag." These hashtags are prefixed by a pound sign ("#") and are used to denote unique people, events, or ideas, making them useful for studying the spread of information online[25,26,27].

To test the sensor hypothesis, we need a sample of individuals with higher network centrality (the "sensor" group) to compare with a sample of randomly-chosen individuals (the "control" group). However, measuring centrality can be a computationally expensive task in large-scale networks like Twitter. Therefore, we use a simplified approach that first randomly selects a set of users for the control group, and then randomly chooses one "friend" of each member of this group to put in the sensor group. This procedure generates a sensor group with higher centrality than the control group because of the "friendship paradox": high-degree individuals are more likely to be connected to a randomly chosen person than low-degree individuals[22,28]. In other words, "your friends have more friends than you do."[29]

In Fig.1a we demonstrate that the sensor group contains more high degree individuals and fewer low degree individuals, and this is true even if we remove duplicates from the sensor group (duplicates occur when the same person is randomly chosen as a friend by multiple individuals in the control group). However, this difference between the sensor and control groups depends on what fraction of the network is sampled. As the fraction increases, there is increasing overlap between the two groups, reducing the difference in their degree distributions (Fig.1b). We derive closed form equations that characterize the expected degree distribution for both the sensor groups (with and without duplicates) and control groups based on the fraction of nodes sampled and an arbitrary known degree distribution for the network as a whole (see SI). Fig.1c,d show that these equations fit the data well for a random sample of 1.25% of all users (500,000 total) on Twitter, confirming our expectation that the sensor group is more central than the control group.

To test whether sensors can provide early warning of a contagious message spreading through the network, suppose $t_i^\alpha$ denotes the time at which a sampled user $i$ first mentions hashtag α (i.e the infection time). We would expect $t_i^\alpha$ to be smaller on average for users belonging to a central sensor group $S$ than for those of a random control group $C$. If we denote $\Delta t^\alpha = \langle t_i^\alpha \rangle_{i \in S} - \langle t_i^\alpha \rangle_{i \in C}$ for hashtag α, the sensor hypothesis is that $\Delta t^\alpha < 0$.

However, note that $\Delta t^\alpha$ depends on the size of the samples in two ways. For small samples, the number of "infected" users (i.e. users mentioning hashtag α) will be scarce, leading to large statistical errors. On the other hand, for big samples, the degree distribution the of control and sensor groups tend to overlap and consequently $\Delta t^\alpha$ approaches 0. Therefore it may be necessary to find an optimal "Goldilocks" sample size that gives statistical power while still preserving the high-centrality characteristic of the sensor group. Fig.2a shows results from a theoretical simulation of an infection[30] spreading in a synthetic network (see SI) while Fig.2b shows an empirical analysis of widely used hashtags in our Twitter database. Both theory and data suggest that there exists an optimal (and moderate) sample size that may perform best for detecting large and significant differences between the sensor and control group resulting from contagious processes.

To analyse the performance of the sensor mechanism, we collected five random control samples of 50,000 users and a random set of their followees of the same size to use as sensors for each one. Focusing on the 32 most widespread hashtags that appear at least 10 times in each control sample, Fig.2c shows that $\Delta t^\alpha$ is negative (i.e., the sensor sample uses the hashtag prior to the control sample) in all but two cases, with a mean for all hashtags of –7.1 days (SEM 1.1 days). In the SI we also show this distribution for a wider range of hashtags, and these all show that $\Delta t^\alpha$ tends to be negative. In other words, the sensor groups provide early warning of the usage of a wide variety of hashtags.

We also hypothesized that comparative monitoring of a sensor group and a control group may help distinguish which hashtags are spreading virally via a contagious process and which are spreading via broadcast. We studied 24 hashtags (Fig.3a) that were "born" during our sample period (they first appeared at least 25 days after the start date of data collection) and then became widely used (they were eventually used more than 20,000 times). Notably, the users using these hashtags tended to be highly connected and many were connected to a giant component, a sign that the hashtags may have spread virally online from user to user (see Fig.3d and SI for more examples).

For each of these hashtag networks, we constructed a random control sample of 5% its size and a similarly-sized sensor sample of their followees to calculate $\Delta t^\alpha$. We then repeated this process 1,000 times to generate a statistical distribution of these observed lead times (Fig.2b). The sensor group led the control group ($\Delta t^\alpha < 0$) 79.9% (SE 1.2%) of the time. However, note that there was considerable variation in lead times, from 20 days to a few hours or no advance warning.

To see how the sensor method works for hashtags that are *not* spreading virally, we generated a null distribution in which we randomly shuffled the time of each hashtag use within the fully observed data, and then measured the resulting difference in the sensor and control group samples, $\Delta_R t^{iz}$. Again, we repeated the procedure 1,000 times to generate a statistical distribution (see SI). The results show that the observed distribution of lead times falls outside the null distribution for 65.4% (SE 1.2%) of the hashtags, suggesting they did, in fact, spread virally (Fig.3a).

Most hashtags also showed a shift forward in their daily and cumulative incidence curves of the sensor group compared to the control one (Fig.3c,d). This shift forward, another sign of virality in itself, could allow for identification of an outbreak in advance, as the sensor's deviation from the trajectory of the control group identifies a process that is spreading through the network, affecting central individuals faster than random ones. Estimating the models each day using all available information up to that day, for #openwebawards users, we find two consecutive days of significant ($p<0.05$) lead time by the sensor group compared to the control group on day 13, a full 15 days before the estimated peak in daily incidence (see SI), and also 15 full days before the control sample reaches the same incidence as the sensor group (See Figure 3c).

Finally, while the sensor mechanism allows us to identify a more central group that can be used to detect contagious outbreaks in advance, it may also allow us to focus on users who have other characteristics that could improve monitoring. In particular, users in the sensor group may be more central because they are more active, and indeed we find this to be true (Fig.4a). On average, users in the sensor group sent 154 tweets (SE 2.8) during the six months they were monitored, while users in the control group tweeted only 55 times (SE 1.0, difference of means *t*

= 36, $p$ < 2.2e-16).   However, we also find that more central users tend to use a greater variety of hashtags, even controlling for activity levels (Fig.4b).

The distribution of the number of users using any one hashtag follows a heavy tailed distribution (see SI) with most hashtags being used by less than a few hundred people and very few reaching the tens of thousands.   Therefore, for most hashtags, the probability of finding sufficient users to perform a significant analysis in a random sample of Twitter is very small. Yet despite the relatively small size of the infected populations, the sensor mechanism we test here seems to anticipate the global spread of information in a wide variety of cases.   And, importantly, it only requires a tiny fraction of the network as a whole to be monitored.

We believe that this method could be applied in a wide variety of contexts in which scholars, policy-makers, and companies are attempting to use "big data" online to predict important phenomena.   For example, the sensor method could be used in conjunction with online search to improve surveillance for potential flu outbreaks[22,2].   By following the online behaviour of a group known to be central in a network (for example, based on e-mail records which could be used to construct a friend sensor group), Google or other companies that monitor flu-related search terms might be able to get high-quality, real-time information about a real-world epidemic with greater lead time, giving public health officials even more time to plan a response. Similarly, policy-makers could monitor global mood patterns[6] to anticipate important changes in public sentiment that may influence economic growth, elections, opposition movements, or even political revolutions[8].   We also conjecture that investors might use these methods to better predict movements in the stock market[7].

How much advance detection would be achieved in other contexts or in populations of different size or composition remains unknown.   Just as we find variation in lead time for different hashtags, we expect that the ability of the sensor method to detect outbreaks early, and how early it might do so, will depend on intrinsic properties of the phenomenon that is spreading; how this is measured; the nature of the population, including the overall prevalence of susceptible or affected individuals; the number of people in the sensor group; the topology of the network (for example, the degree distribution and its variance, or other structural attributes)[23]; and other factors, such as whether the outbreak modifies the structure of the network as it spreads (for example, by affecting the tendency of any two individuals to remain connected after the information is transmitted). In future work, we hope to explore these factors in order to generalize the conditions under which friends can be used effectively as sensors to detect contagious outbreaks.

**Acknowledgements** This research was supported by a grant from the National Institute on Aging (P-01 AG-031093) and by a Pioneer Grant from the Robert Wood Johnson Foundation.

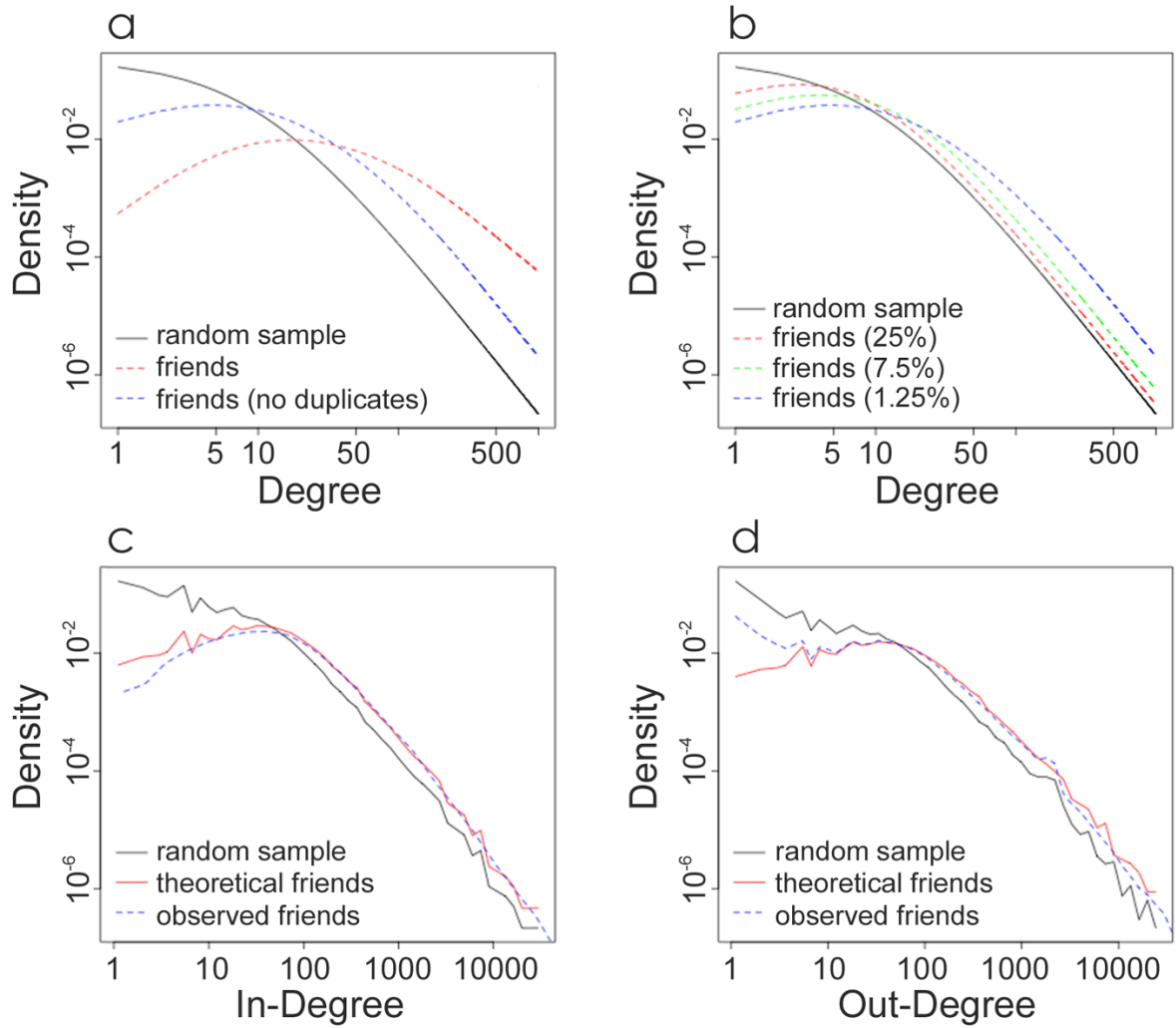

**Figure 1**: Twitter exhibits the "friendship paradox": a) Expected degree distributions for a 1.25% random sample of the Twitter network (black line), friends of this randomly chosen group (red line), and the same friends group with duplicates removed (blue line); b) Larger samples of friends show a smaller difference in degree distribution from the overall network (black = overall network, red = 25% sample, green = 7.5% sample, blue = 1.25%); c) and d) Respectively, In-degree (follower) and out-degree (followee) distribution of a random sample of 500,000 users, 1.25% of Twitter's users (the "control" group, black line) and the theoretical (red line) and observed (blue line) in-degree and out-degree distributions of their friends (the "sensor" group) with duplicates from the friends group removed.

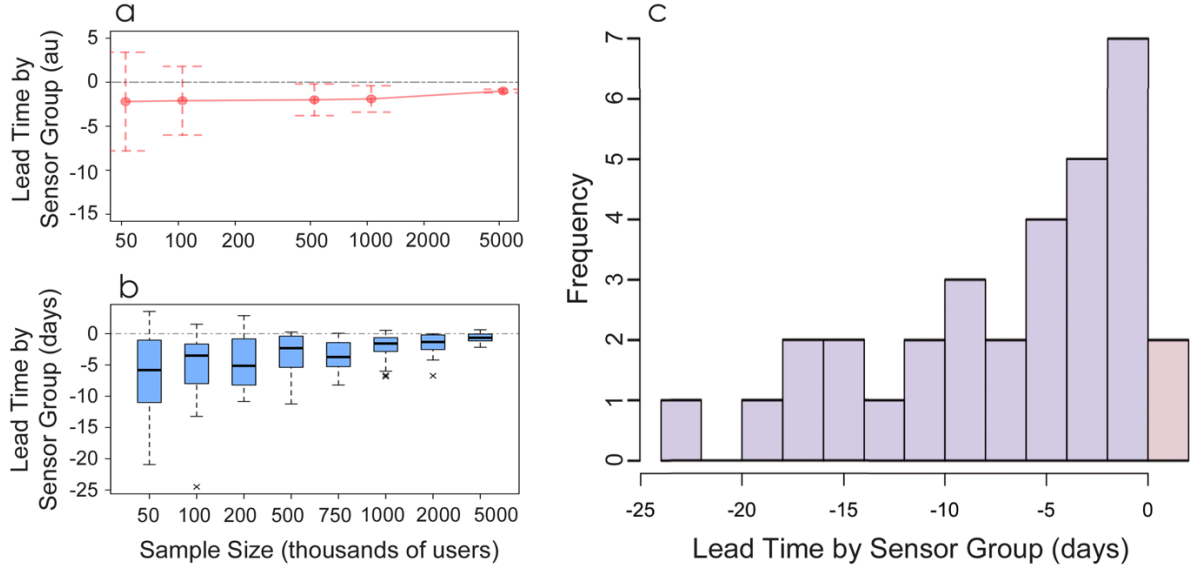

**Figure 2**: Friends as sensors yield early detection of the use of hashtags. a) Measures of lead times based on simulations of an infection spreading through a network with infection probability $\lambda=0.1$ and recovery probability $\gamma=0.01$ on a Barabasi-Albert random network with tail exponent $\beta>3$ show that a sensor group tends to provide earlier warning than a randomly-chosen control group in smaller samples, but decreasing sampling variation in larger sample sizes means that the statistical likelihood of providing early warning is maximized in moderately-sized samples. b) Observed results for hashtags on Twitter used by 1% of the individuals using a hashtag of each sample. c) Average lead time of first usage of each hashtag in the sensor group vs. the control group for all hashtags used by at least 10 users in each of 5 random samples of 50,000 random users.

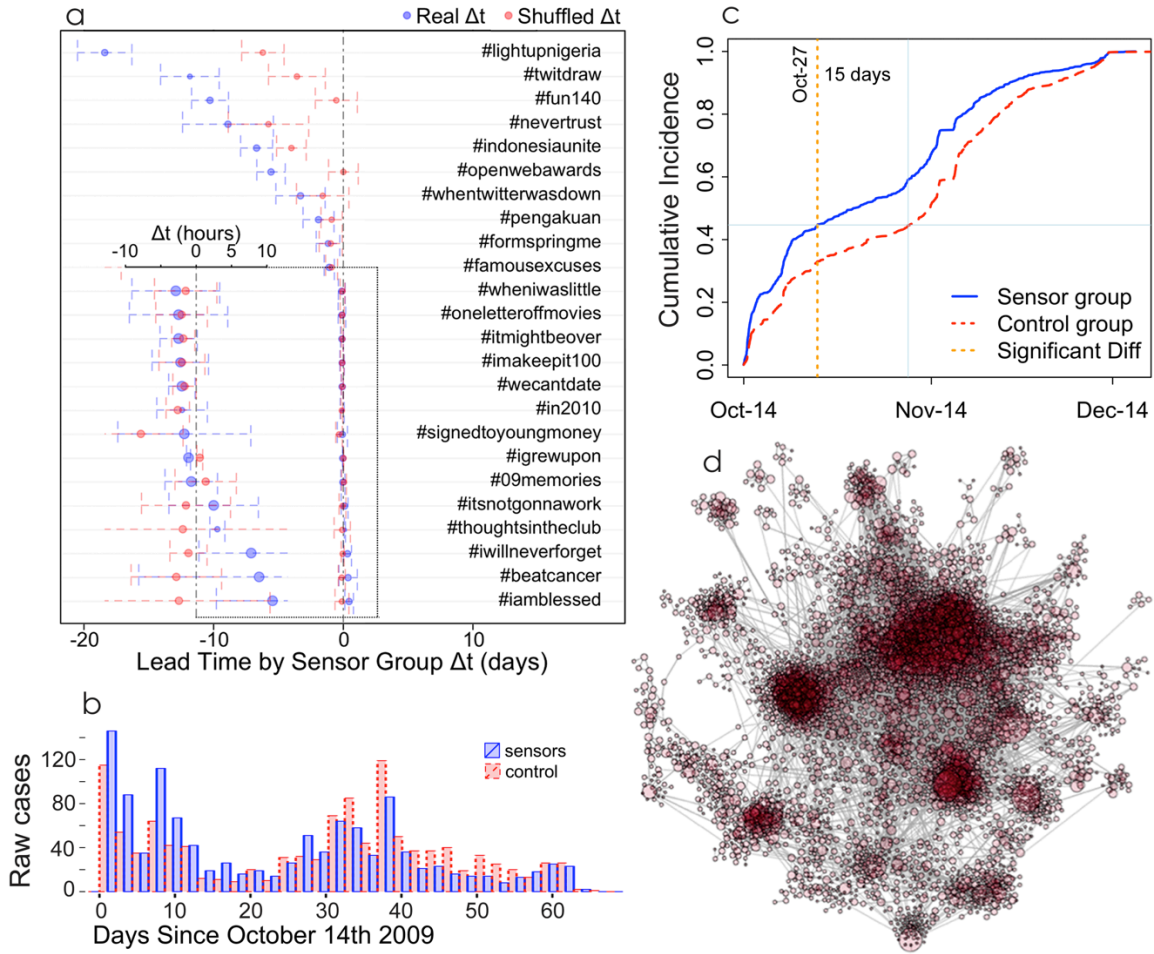

**Figure 3**: Signs of virality in hashtag usage.   a) The average lead for the 24 most-used hashtags time across 1,000 trials of the sensor group (in blue) vs. the same calculated lead time when all times of first usage are randomly shuffled (in red).   Vertical bars are SEM.; b) daily incidence and c) cumulative daily incidence for the hashtag #openwebawards show a shift forward in the S-shaped epidemic curve and a burst in the sensor group relative to the control group that could be used to predict the outbreak of this hashtag on the 13[th] day (the first day on which, using all available information up to that day , there is a significant difference between the sensor and control groups with p-value < 0.05), 15 days before the control group reaches the same cumulative incidence and before the estimated peak in daily incidence; d) greatest connected component of the follower network of users using the #openwebawards hashtag shows that many users are connected in a large component.

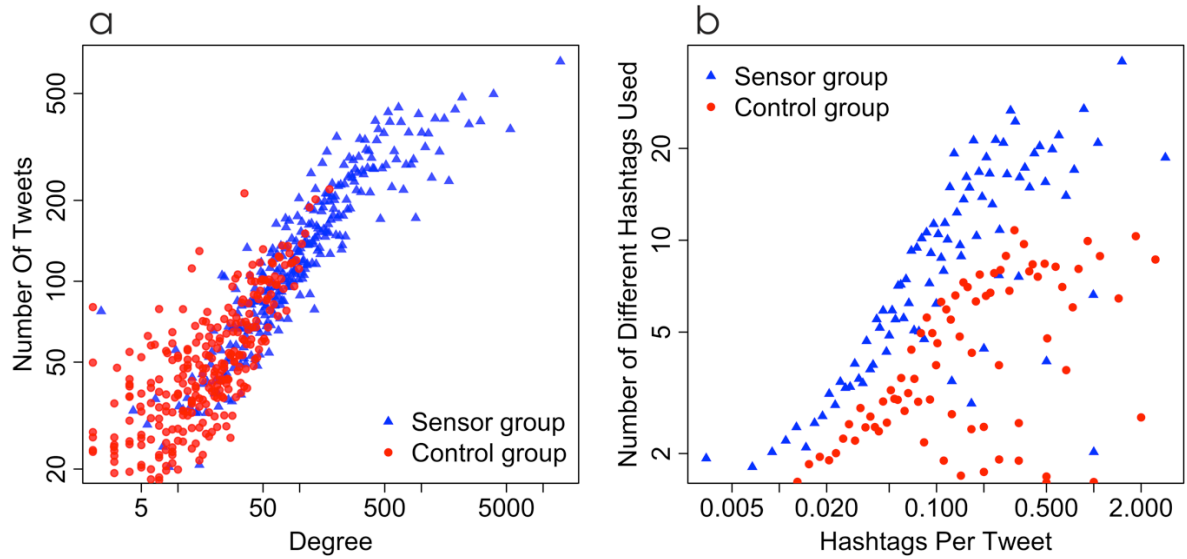

**Figure 4**. Users in the sensor group (blue) are more active (a) and also use a wider variety of hashtags (b) than those in the control group (red), even controlling for activity. These attributes both contribute to early warning provided by the sensor group's structural position.

# Supplemental Information for

# Using Friends as Sensors to Detect Global-Scale Contagious Outbreaks

Manuel Garcia-Herranz[1], Esteban Moro Egido[2,3,4], Manuel Cebrian[5,6,7], Nicholas A. Christakis[8,9], James H. Fowler[10,11,*]

[1] *Escuela Politécnica Superior, Universidad Autónoma de Madrid, Madrid 28049, Spain*
[2] *Departamento de Matematicas & GISC, Universidad Carlos III de Madrid, 28911 Leganés, Spain*
[3] *Instituto de Ingeniería del Conocimiento, Universidad Autónoma de Madrid, Madrid 28049, Spain*
[4] *Instituto de Ciencias Matemáticas CSIC-UAM-UCM-UC3M (ICMAT), 28049, Madrid, Spain*
[5] *Computer Science & Engineering Department, University of California, San Diego, CA 92093, USA*
[6] *Media Laboratory, Massachusetts Institute of Technology, Cambridge, MA 02139, USA*
[7] *National Information and Communications Technology Australia, Melbourne, Victoria 3010, Australia*
[8] *Department of Sociology, Harvard University, Cambridge, MA 02138, USA*
[9] *Department of Medicine, Harvard Medical School, Boston, MA 02115, USA*
[10] *Medical Genetics Division, School of Medicine, University of California, San Diego, CA 92103, USA*
[11] *Political Science Department, University of California, San Diego, CA 92103, USA*

*Corresponding Author: JHF, jhfowler@ucsd.edu

## Table of Contents

# An Analytic Elaboration of the Friendship Paradox

The "Friendship Paradox" was first elaborated by Feld (1991) who proved that for any arbitrary network the average number of friends of friends is greater than the average number of friends ("Your Friends Have More Friends Than You Do").

Mathematically, we can define $\mu$ as the average number of connections per vertex in a network with $V$ vertices and $E$ edges:

$$\mu = \frac{\sum_{v \in V} k_v}{|V|} = \frac{2|E|}{|V|} \quad (1)$$

where each vertex $v \in V$ has $k_v$ connections.

Using the same notation, we can derive the number of two-degree connections $\rho$ (the friends of friends). Since the degree $k_v$ of a vertex $v$ is counted $k_v$ times (if a person has 5 friends, each of them will be added 5 times to the total number of friends of friends), there are $\sum_{v \in V} k_v^2$ two-degree connections, and we can divide this by the total number of direct connections $\sum_{v \in V} k_v$ to get the average number of two-degree connections $\rho$:

$$\rho = \frac{\sum_{v \in V} k_v^2}{\sum_{v \in V} k_v} \quad (2)$$

Using a property of the variance $\sigma^2$ in the distribution of $k_v$, note that:

$$\sigma^2 = E\left[(k-\mu)^2\right] = E\left[k^2\right] - \left(E[k]\right)^2 = \frac{\sum_{v \in V} k_v^2}{|V|} - \frac{\left(\sum_{v \in V} k_v\right)^2}{|V|^2} \quad (3)$$

From this, it is easy to show that:

$$\rho = \frac{\sum_{v \in V} k_v^2}{\sum_{v \in V} k_v} = \mu + \frac{\sigma^2}{\mu} \quad (4)$$

Therefore $\rho$ is greater than $\mu$ ($\frac{\sigma^2}{\mu}$ greater, to be precise) and this difference thus increases with the variance $\sigma^2$ of the degree distribution in the network.

Equation (2) implies that the expected distribution of two-degree connections $Q(k)$ will be a function of the degree distribution $P(k)$. Since each vertex of degree $k$ is connected to $k$ other vertices, each vertex's degree is counted $k$ times in the distribution and therefore there will be $k$ vertices of degree $k$, leading to the following distribution of two-degree connections:

$$Q(k) = \frac{kP(k)}{\mu} \quad (5)$$

Since we want to sample only a part of the network of monitoring, we need to know the dregree distribution we can expect to find for a random sample of a given size. If we sample a portion of the network ($0 \leq \gamma \leq 1$), the degree distribution $P(k)$ will remain the same since every person has the same probability $\gamma$ of being chosen. However, the expected distribution of the two-degree connections for the "friends" sample will differ since a person with more friends has more chances of having a friend being chosen in the random sample and therefore appearing in the friends sample.

The probability of not chosing a person with $k$ friends is equal to the probability of not chosing any of thir $k$ friends $(1-\gamma)^k$ and thus their probability of appearing in the friends sample is $1-(1-\gamma)^k$. Drawing on Equation (5), we can use these probabilities to derive the degree distribution of a random sample and a corresponding friends sample for a portion $\gamma$ of the population:

$$\tilde{Q}(k) = \frac{k\left[1-(1-\gamma)^k\right]P(k)}{A_\gamma} \quad (6)$$

Where $A_\gamma$ is a normalization factor.

For the complete network ($\gamma = 1$), Equation (6) simplifies to Equation (5), and both of these demonstrate the friendship paradox, $Q(k) > P(k)$ for large $k$. This effect is amplified for small $\gamma$ since it maximizes the multiplying effect of counting $k$ times each vertex with degree $k$ (see Figure S1). Thus, a sensor group composed of the friends of a randomly-sampled control group will tend to have a larger number of high-degree vertices than the control group, itself.

One issue with constructing a sensor group is that people with many friends may be friends of several persons of the randomly-sampled control group and therefore may be counted several times as members of the sensor group. In fact, a person with $k$ friends will be counted $k$ times when $\gamma = 1$. However, we remove such duplicates when constructing the sensor sample, so that each person is counted only once. This means that the degree distribution of the sensor group will not be multiplied by $k$:

$$\tilde{Q}(k) = \frac{\left[1-(1-\gamma)^k\right]P(k)}{A_\gamma} \quad (7)$$

Notice that this sensor group with duplicates removed also exhibits higher centrality than the network as a whole since $\left(1-(1-\gamma)^k\right)$ increases with $k$. However, like the friends degree distribution described in equation (6), the difference between the sensor group $\tilde{Q}(k)$ and the control group $P(k)$ disappears as $\gamma$ increases (see Figure S1). Thus, the centrality advantage of the sensor group declines as the size of the sample increases.

## The Twitter Data

We used the data gathered by Kwak et al.[1], collected between the 1st of June 2009 and the end of December 2009, that represents a near-complete graph collected by snowball-sampling the online social networking site, Twitter. Most (active) SNS users are connected to the giant connected component to which Paris Hilton belongs (a user with over one million followers that was used for the snowball sample). Additionally all users mentioning trending topics from June to August were also crawled. Summarizing, the sample includes:

1. those users connected to the giant connected component (no matter whether following or followed)

2. those who mentioned trending topics.

The data includes:
- 476,553,560 tweets
- 66,935,466 tweets using a hashtag (a key term prefixed by '#')
- 4,093,624 different hashtags
- 40,171,624 registered users
- 1,620,896 registered users using at least once a hashtag
- 1,468,365,182 follow relationships
- 531,703,974 reciprocal follow relationships (i.e. 265,851,987 bidirectional links)

The popularity of different hashtags, measured as the number of different people using each of them at some point along this period, approximately follows a power law. In addition, Figure S2 shows a global increase in the number of tweets per day with a seasonal peak in summer as well as a weekly pattern in the number of hashtags used per day.

In Figure S3 we show the popularity and start date for all hashtags appearing for the first time after June 25th, 2009. Since the start date for data collection is June 1st, 2009, and we did not observe these hashtags in the first 25 days, we assume that we have recorded their "birth" in the Twittersphere and therefore their possible initial outbreak.

We focus our analysis on the most used hashtags to analyze their virality in order to identify big local propagation communities. Figure S4 shows the relative size of the greatest components for each network made up of all users using a particular hashtag, and compares it to the total number of users who used the hashtag.

## Sensor Performance in a Simulated Infection Model

To test the theoretical behavior of the sensor mechanism in an environment with a known spreading process, we created a simple program in R to generate a synthetic Barabasi-Albert random undirected network with tail exponent $\beta>3$ over which we simulated 10 cascades (each beginning at a different vertex of the network) and an infection-recovery process with infection probability $\lambda=0.1$ and recovery probability $\gamma=0.01$:

```
library(igraph)

# generate the network
g <- barabasi.game(50000,power=1,m=5,directed=F)
mm <- data.frame(get.edgelist(g))+1 #network edges
colnames(mm) <- c("id1","id2")

tend = 10000   # we simulate the model for 10000 time units

ntot <- max(c(mm$id1,mm$id2))

time <- rep(tend,ntot)    #infection times
infected <- rep(0,ntot)   #infected nodes (0=S, 1=I, 2=R)

ncascades <- 10     #number of cascades
lambda <- 0.1       #infection probability
gamma <- 0.01       #recovery probability

for(k in 1:ncascades){  # simulate 10 casacades
  i0 <- sample(1:ntot,1)  # select a random seed for the cascade
  time[i0] <- 0           # ...that gets initialy infected
  infected[i0] <- 1

  ni <- c(1)   # 1 node infected (just the seed)
  nr <- c(0)   # no nodes recovery yet

  for(i in 1:tend){  # begin the cascade
     #infecteds get recoverd with probability gamma
        infected[infected==1 & (runif(ntot) < gamma) ] = 2

        ii <- which(infected==1) # get the list of still infected
```

```
        nb <- unique(c(mm[is.element(mm$id1,ii),2],mm[is.element(mm$id2,ii),1]))
        #neighbors of the infected nodes

        # Sucsceptible neighbors get infected with prob. lambda
        nb <- nb[runif(length(nb)) < lambda]
        nb <- nb[infected[nb]==0]
        infected[nb] <- 1

        time[nb] <- I  # record time of infection

        ni <- c(ni,sum(infected==1)) # update infected list
        nr <- c(nr,sum(infected==2)) # update recovery list
        if(ni[i]==0) break  # stop if there are no more infected
    }
}
```

We then used data generated by these simulations to collect 350 random samples of 7 different sizes (0.124% of the network to 12.5% of the network), 50 of each, and a similarly sized sample of its neighbors to use as sensor:

```
# function to get the neighbors of a node
neighbors <- function(id){
    unique(c(mm[mm$id1==id,2],mm[mm$id2==id,1]))
}

samples <- c(62,125,312,625,1250,2500,6250) # sample sizes

final <- c()
for (samp in 1:length(samples)){ # for eac sample size...
    dtc <-c()  # initialize control times
    dts<-c()   # initialize sensor times

    for(i in 1:50){ # repeat 50 times

        # get a control sample
        control <- sample(1:ntot,samples[samp])

        # and a sensor sample of its neighbors \wo repetition
        sensor <- sample(unique(unlist(lapply(control, neighbors))),samples[samp])

        # get the times of infection of the infected nodes
        tcontrol <- time[control]
        tsensor <- time[sensor]
        tcontrol <- tcontrol[tcontrol!=tend]
        tsensor <- tsensor[tsensor!=tend]

        # get the mean time of infection of controls
        dtc <- c(dtc,mean(tcontrol))
        #and sensors
        dts <- c(dts,mean(tsensor))
    }

    finalmean <-mean(dts-dtc) # mean Delta t
    finalesem <-sd(dts-dtc)/sqrt(length(dts)) # SEM Delta t

    Save the mean and SEM Delta t of this sample size
    final<-rbind(final,c(as.character(sample[samp]), as.numeric(finalmean),
    as.numeric(finalesem)))
}
finaldf <- data.frame(final)
colnames(finaldf)<-c("sample","mean","sem")
```

The results can be seen in Figure 2a of the main text, showing that as the sample size grows the SEM decreases and the mean Δ$t$ approaches 0, suggesting that middle sized samples are the best choice for detecting significant lead times.

# Sensor Performance in Real Data

We analyzed random control samples of various sizes (50K, 100K, 250K, 500K, 1M, 1.5M, 2M and 5M users), and compared each one against 30 random sensor samples of the same size. In these sensor samples we ensured that no user appeared more than once within a single sensor sample.

In order to detect a viral process with the sensor mechanism two factors are crucial. First, the sensor sample should be more central than the control one. Second, the viral outbreak is large enough to be detected. We study viral outbreaks in retrospect (once the spreading process is finished), focusing our attention on those that affected the largest number of nodes in the network.

In Figure S5 all hashtags used by more than 0.01% of each sample's users (or about 0.25% of all hashtag users) have been plotted and compared to the theoretical infection model, finding that the mean $\Delta t^{\alpha}$ and variance approach 0 as the sample size grows, as predicted by the model. Figure S6 shows the same thing for hashtags used by more than 0.04% of the sample (or about 1% of hashtag users). Notice that in both cases, small sample sizes are not large enough for lead times ($\Delta t^{\alpha}$) to be reliably less than zero (due to sampling variation), while in large samples the control and sensor groups tend to overlap, causing the mean $\Delta t^{\alpha}$ to approach 0 (see Figure 2a,b in the main text).

Many of the important events of 2009 appearing on Twitter show a lead time for the sensor group. They also show a relationship with the control group by sample size similar to that of the infection model. For example, Figure S7 shows hashtags for:

- *#h1n1*: Current level of the novel H1N1 influenza pandemic alert raised from phase 5 to 6, June 11
- *#indonesiaunite*: Sucide bombers hit two hotels in the center of Jakarta, July 17
- *#cop15*: UN Climate Change Conference 2009, December 7
- *#iran*: Protests following 2009 Iranian presidential election, June 13
- *#michaeljackson*: Michael Jackson's death, June 25.
- *#forasarney*: Social movement demanding the departure of Senator Jose Sarney of their duties in the National Congress of Brazil, June 17.

Figure S8 shows that viral hashtags that grow in usage and spread from person to person over time like *#mobsterworld* and those used in response to an exogenous event like *#indonesiaunite* show negative $\Delta t^{\alpha}$, suggesting that the sensor method elaborated here may be a reliable way to predict their widespread usage. In contrast, a few hashtags like *#health* yield a positive $\Delta t^{\alpha}$ (when the sensor group lags the control group). There is variation in the amount of lead time provided by the sensor group. Hashtags like *#beatcancer*, a one day campaign granting donations for each tweet using it, spread extremely quickly through the Twitter network. As a result, they display an almost flat curve around zero and no variance, suggesting that, in these cases, the sensor mechanism is not sensitive enough to predict an epidemic in advance.

Additional examples of viral and nonviral hashtags can be found in Figures S9 and S10. We also show the network of users using a given hashtag and the cumulative distribution for the sensor and control groups for several example hashtags in Figures S11-S14. Notice that many of these show that the cumulative distribution of usage in the sensor group is shifted to the left, a sign that the center of the network is being infected before the network as a whole, on average.

# Using the Sensor Method with a Small Set of Samples

Here, we study the use of multiple sensor and control samples for detecting contagious outbreaks. We choose 5 random samples of 50,000 Twitter users (to be used as control) with their complete set of followees (from which to obtain the sensor samples), calculating their in and out degree and the times at which they mention any hashtag in a tweet. Obtaining a complete list of followers from a random sample over the whole Twittersphere is a costly process, especially for large samples. Therefore we reduced our statistical analysis to 5 samples and followed the same methodology we later repeated with 1,000 samples when we did not need to tackle Twitter as a whole.

We trim the control samples to those users that have mentioned a hashtag at least once. Since only 4% of Twitter users have ever used a hashtag, our control samples are trimed to about 2,000 users and the followees samples to about 150,000 users.

For each control we randomly select a similarly-sized sensor sample (without duplicates) from its followees. We then remove all hashtags that have appeared before June 20$^{th}$ 2009 (20 days after our first records) reducing our analysis to those hashtags that were very likely "born" during our sample period. For each of the hashtags we calculate the mean first time of use in the control and sensor samples. We then calculate the mean control time minus the mean sensor time for those hashtags used by at least 10 individuals in 1 or more of 5 samples.

Notice that the total population of these hashtags can be described with the following probability distribution:

$$P(\alpha, n_s, s) = \sum_{i=s}^{n_s} \left( \binom{n_s}{i} P(\alpha)^i (1-P(\alpha))^{n_s - i} \right) \qquad (8)$$

Where $n_s$ is the total number of samples (i.e. 5) and $s$ is the minimum number of samples (i.e. 1) in which we require the hashtag to appear in at least $x_s$ users (i.e. 10). Therefore it is the probability of appearing in $s$ samples and not appearing in the other $n_s - s$ samples multiplied by the number of posible combinations of $n_s$ samples taken in groups of $s$, plus the same probabilities for every $I$ samples where $s > i \geq n_s$.

Now, the probability of finding $x_s$ or more users of a hashtag $\alpha$ in a random sample of $S$ users drawn from a total population of $N$ users is defined through the hypergeometric function, which indicates the probability of finding exactly $x_s$ individuals of the $X_\alpha$ users of hashtag α in a random sample of $S$ users drawn from a population of $N$ individuals:

$$P(n_s; N, X_\alpha, S) = \frac{\binom{X_\alpha}{n_s}\binom{N-X_\alpha}{S-n_s}}{\binom{N}{S}} \qquad (9)$$

The probaility of finding $x_s$ or more individuals is simply the cumulative distribution of Equation (9):

$$D(n_s; N, X_\alpha, S) = \sum_{k=n_s}^{X_\alpha} P(k; N, X_\alpha, S) \qquad (10)$$

while the cumulative distribution for not finding $x_s - 1$ or less individuals is:

$$P(\alpha) \equiv D(n_s; N, X_\alpha, S) = 1 - \sum_{k=0}^{n_s - 1} P(k; N, X_\alpha, S) \qquad (11)$$

Where $P(\alpha)$ is the probability for hashtag α to have at least $x_S$ users in a random sample of *S* users drawn from a population of *N* individuals.

Complementing equation (8), equation (11) allows us to estimate the number of total hashtag users for hashtags with 10 or more users appearing in 1 or more of the 5 samples, 2 or more, and so on.   The distribution of the total number of users of hashtags appearing in each of these cases and the distribution of their mean $\Delta t$ can be seen in Figure 2c of the main text and Figure S15.

Forcing a hashtag to have representatives in all 5 samples restricts our analysis to the most widespread hashtags (32 in our case), while lowering this restriction to hashtags that have enough representatives in at least one of the samples opens up the analysis to more hashtags (134 in our case) even though they do not necessarily have a global spread as wide as the former.   In every analysis (i.e. hashtags having enough representatives in 1 or more of the 5 samples, 2 or more, 3 or more, 4 or more and in all of them) more than 70% of the hashtags show a negative $\Delta t^{\alpha}$ (in 80.55% in average 7.27% sd)

Nevertheless, the results show that raising the threshold number of samples required for detection of the global spread of a hashtag reduces noise and yields a $\Delta t < 0$ for nearly all hashtags, providing advance warning of their outbreak. In other words, the wider audience a hashtag has reached, the greater probability that the sensor mechanism will detect it in advance.

## Using the Sensor Method with Hashtag Networks

Given the comparatively low number of hashtag uses on Twitter (about 14% of the tweets and only 4% of the users use hashtags) and the high diversity of hashtags (4 million different hashtags for 66 million tweets displaying a hashtag), the probability of finding enough users of a hashtag for statistical analysis in a small random sample of Twitter is very small.

Therefore, to analyze particular hashtags, we focused on the complete network of their users, composed only of those people that have used the hashtag in at least one tweet and their follow relations. This allows us to individually analyze the dynamics of well known hashtags as well as to try to identify smaller networks through which a viral process has spread.

We selected all hashtags that were used more than 20,000 times on Twitter, identified their users and their follow relations to construct, for each of them, its hashtag network.
For each of these hashtag networks, we selected a random control sample of 5% its size and a similarly-sized sensor sample of their followees to calculate $\Delta t^{\alpha}$.   We then repeated this process 1,000 times to generate a statistical distribution of these observed lead times (Fig.2b of the main text).    The sensor group led the control group ( $\Delta t^{\alpha} < 0$ ) 79.9% (SE 1.2%) of the time (52.9%, SE 1.12% for $\left[\Delta t^{\alpha}_{p-value=0.05}\right] < 0$ ).   However, note that there was considerable variation in lead times, from 20 days to a few hours or no advance warning.

Moreover, one could use the difference in the sensor and control group to determine whether or not a viral process is under way.    Estimating the models each day using all available information up to that day, for a control and a sensor sample of #openwebawards users, we find a significant lead time (p<0.05) on day 13, a full 15 days before the estimated peak in daily incidence, and also 15 full days before the control sample reaches the same incidence as the sensor group has on day 13 (See Figure 3c in the main text and right figures of Figures S11-S14).

While these results are promising, a $\Delta t^{\alpha} < 0$ does not necessarily indicate a viral process.    Consider a non-viral hashtag with fixed probability of use λ.    A user tweeting at a rate of 10 tweets per hour will use a hashtag with probability 10λ in the first hour, 10 times bigger than that of a user tweeting at a rate of 1 tweet per hour.   Due to the correlation between degree and tweet frequency (see Figure 4a in the main text) a $\Delta t^\alpha < 0$ could be observed for a hashtag not because of its virality but because of the higher tweeting rates of users in the sensor group.

To see how the sensor method works for hashtags that are not spreading virally, we generated a null distribution in which we randomly shuffled the time of each hashtag use within the fully observed data, and then measured the resulting difference in the sensor and control group samples, $\Delta_R t^\alpha$. This method preserves any effect that correlation between degree and activity level might have on lead times. Again, we repeated the procedure 1,000 times to generate a statistical distribution (see SI). The results show that the observed distribution of lead times falls outside the null distribution for 65.4% (SE 1.2%) of the hashtags (25%, SE 0.95% for $\left\lceil \Delta t^\alpha_{p-value=0.05} \right\rceil < \left\lfloor \Delta_R t^\alpha_{p-value=0.05} \right\rfloor$), suggesting they did, in fact, spread virally.

Notice that this method gives us greater confidence that the sensor method detects viral processes, but it also suggests that the non-viral component of Δ*t* could help the sensor mechanism work better than expected, since sensors not only get reached sooner by an epidemic (due to their centrality) but also react earlier (due to their higher tweeting rates).

## Differences in Sensor and Control Characteristics That Also Affect Propagation

By construction, control and sensor groups differ in the local structure of their networks. This is reflected in properties such as degree, betweenness or *k*-coreness. For instance, in the #openwebawards user network, for 30 control/sensor samples of 10% its user base, sensor users have a mean degree of 24.6 (SEM 1.4), 6.6 times greater in average (SEM 0.6) than the 4.5 (SEM 0.6) mean degree of control users, and a mean betweenness centrality 6.2 times greater too (SEM 0.4), 17,615.9 (SEM 492.9) compared to the 3,119.7 (SEM 192.5) of the controls. For 340 random control and sensor samples of 34 different hashtag networks (from 4,645 to 108,073 users) in only 3 cases did the sensor group have a lower mean degree than the control group. Nevertheless, in all of them the sensor group had a higher mean betweenness than the control one (see Figure S16).

Nevertheless, this is not the only factor affecting how much a user may contribute to a contagious outbreak. They may also exhibit different patterns of communication that contribute to the spread of information from person to person to person.

From a 1,000,000 random user sample we identified a control group of 36,499 users using at least one hashtag and a sensor group of the same size composed of the control group's followees. After removing from users from the control group that are present in the sensor group, we end up with 16,332 users in the control group.

Figure 4a of the main text shows how the sensor and control groups differ by degree. The mean number of followers and followees are 25 and 28 in the control group, respectively, but they are 422 and 243 in the sensor group. These numbers reiterate the main result from Figure 1 that sensor samples tend to be much more central than control samples. But there is also a large difference in communication behavior. The sensor group tweets more (152 tweets per user compared to the control group's 55) and also uses more hashtags per tweet (26 compared to the control group's 9). Both of these comparisons show that the sensor group is about three times more active than the control group, but Figure 4a shows that this is a trivial consequence of the relationship between number of tweets and number of connections. In both the control group and the sensor group, the better connected you are, the more messages you send.

Although the *quantity* of communication is the same for the sensor and control group once we control for differences in connectivity, we find a difference in the *quality* of these messages net of the difference in activity levels. Since hashtags are used to denote topics, we can study the extent to which users participate in many different topics or just a few.

Analyzing the number of different hashtags used per tweet and the number of unique hashtag uses, we find a different trend between the control and sensor samples (see Figure 4b in the main text). Sensors use each different hashtag 2.53 times on average (95 percent confidence interval: 2.23 2.83), slightly more than control users who use each hashtag on average 2.15 times (95 percent confidence interval: 2.03 2.28).   Nevertheless, as we have shown, sensors tweet significantly more than controls and when controling for the *frequency* of hashtag usage (hashtags per tweet), we find that users in the sensor group tend to use a significantly wider *variety* of hashtags than control users (i.e. they get involved in a larger number of topics), 96.85 different hashtags per tweet (95 percent CI: 93.51 to 100.20) compared to 38.66 different hashtags per tweet for the controls (95 percent CI: 37.47 to 39.85).

This fact could be of special importance when using the sensor mechanism to anticipate hashtag epidemics as sensors act not only as social hubs (by having more connections) but also as faster responders (by tweeting more) and as information hubs (by being involved in more topics).

# Figures

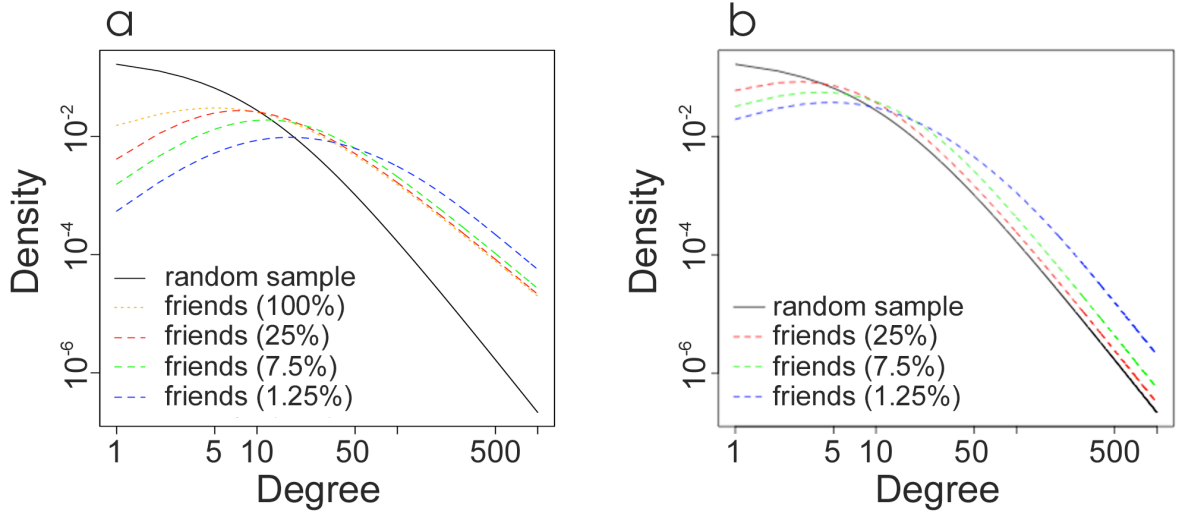

**Figure S1**. Larger samples of friends (as a percentage of the total number of users in he network) show a smaller difference in degree distribution from the overall network both for a) groups of friends with duplicates (e.g. a friend of two users is counted twice) and b) groups of friends without duplicates (a friend of several users is counted only once)

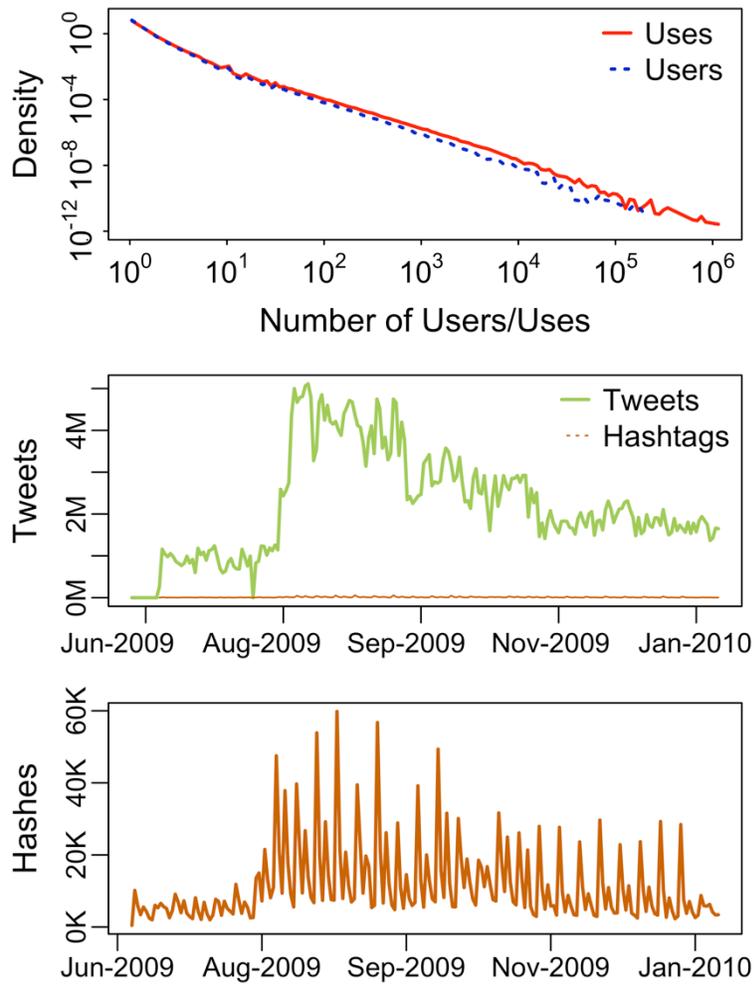

**Figure S2**. Popularity distribution of all hashtags appearing between June 2009 and January 2010 measured as the number of users of each hashtag. This distribution approximately follows a power law. The bottom plots show the total number of tweets and hashtag uses per day.

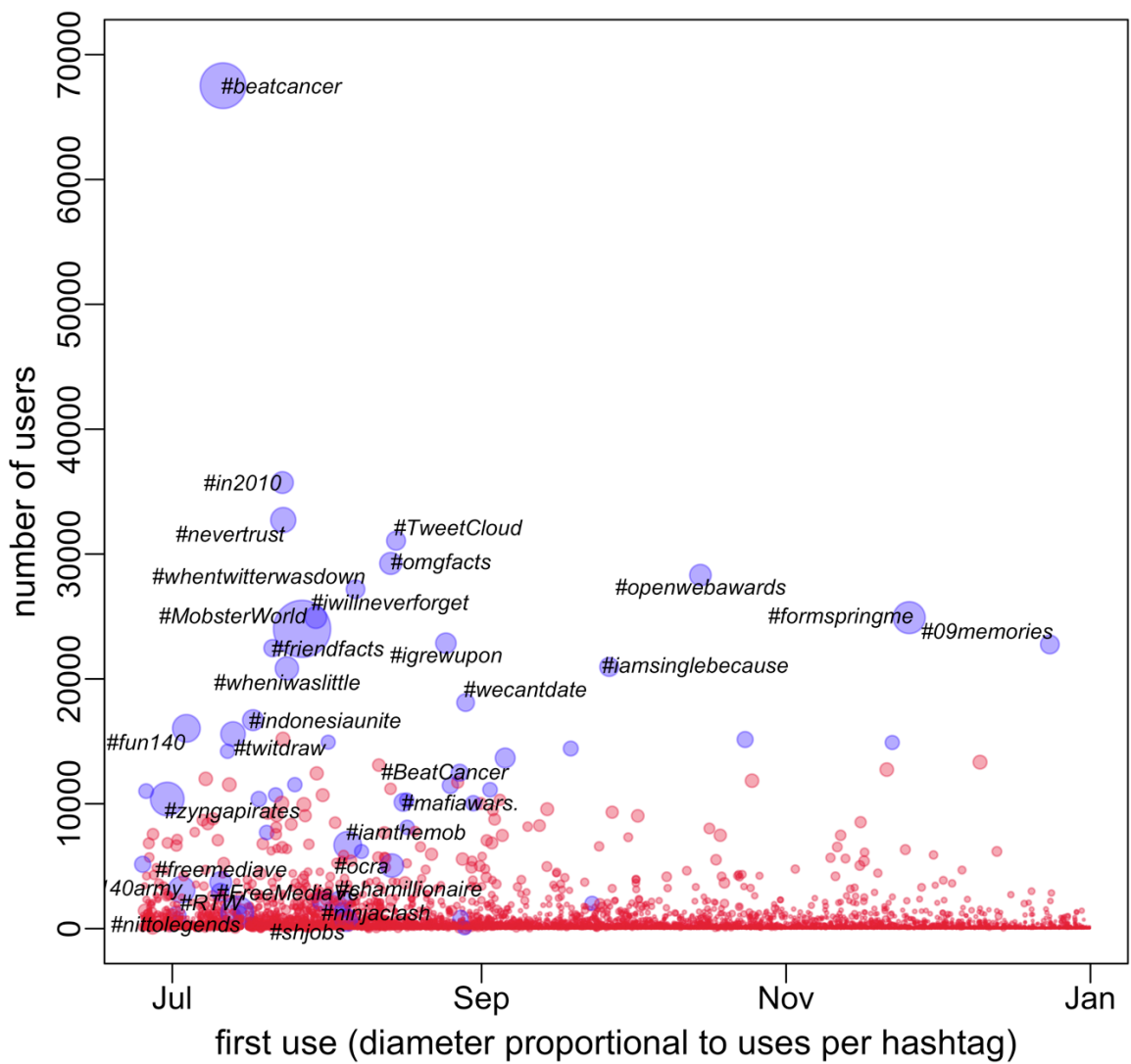

**Figure S3.** Appearance time and number of users of all hashtags appearing for the first time after June 25th. The diameter is proportional to the number of times the hashtag has been used. In blue are all hashtags used more than 20,000 times.

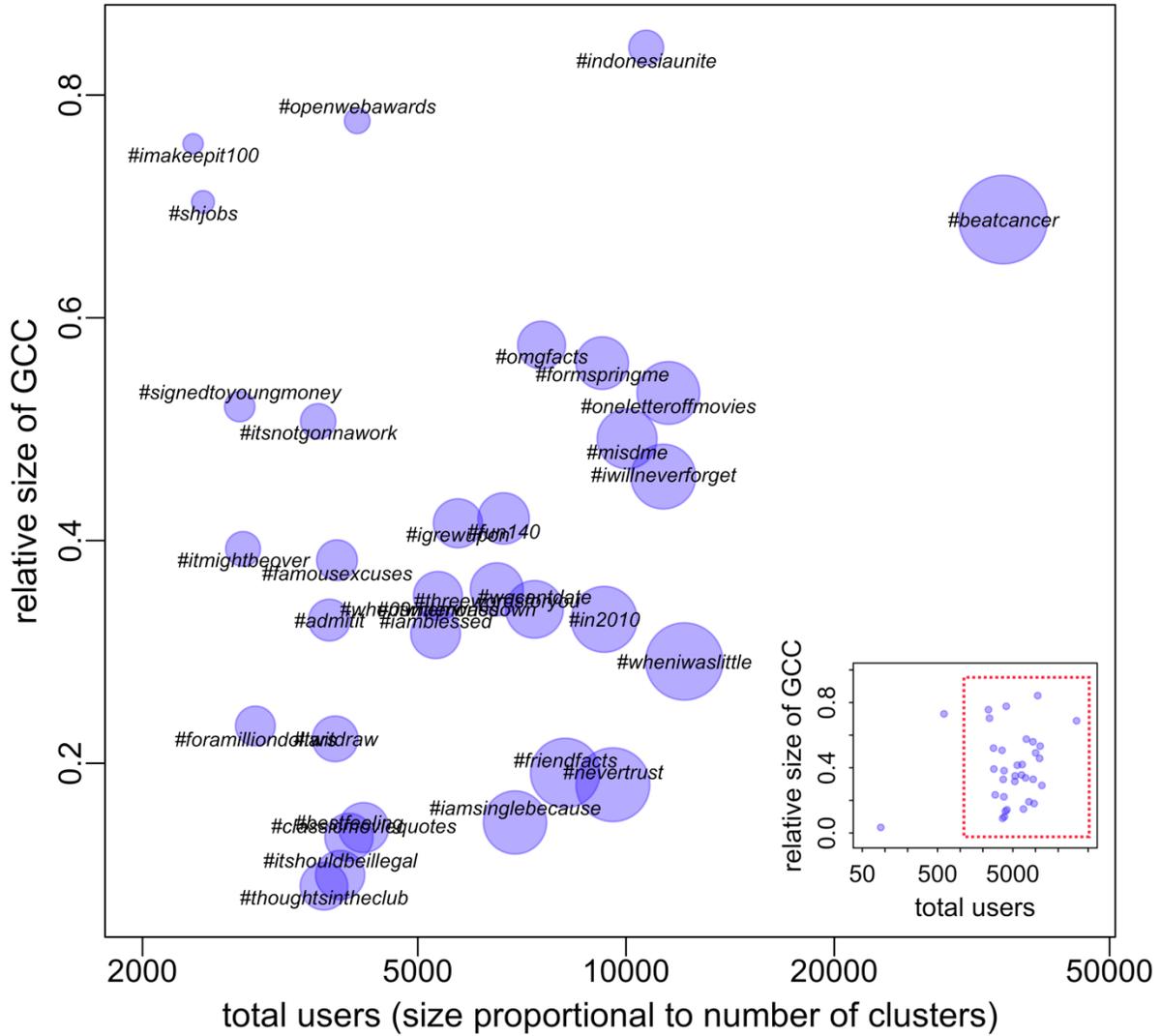

**Figure S4**. Number of nodes in each hashtag network (composed of all users of a particular hashtag and their follow relations to others using the hashtag) and relative size of its greatest component for all hashtags used more than 20,000 times (those in blue in Figure S3). The size of the nodes is proportional to the number of components of the network. Most networks show a large greatest component and many small isolated components.

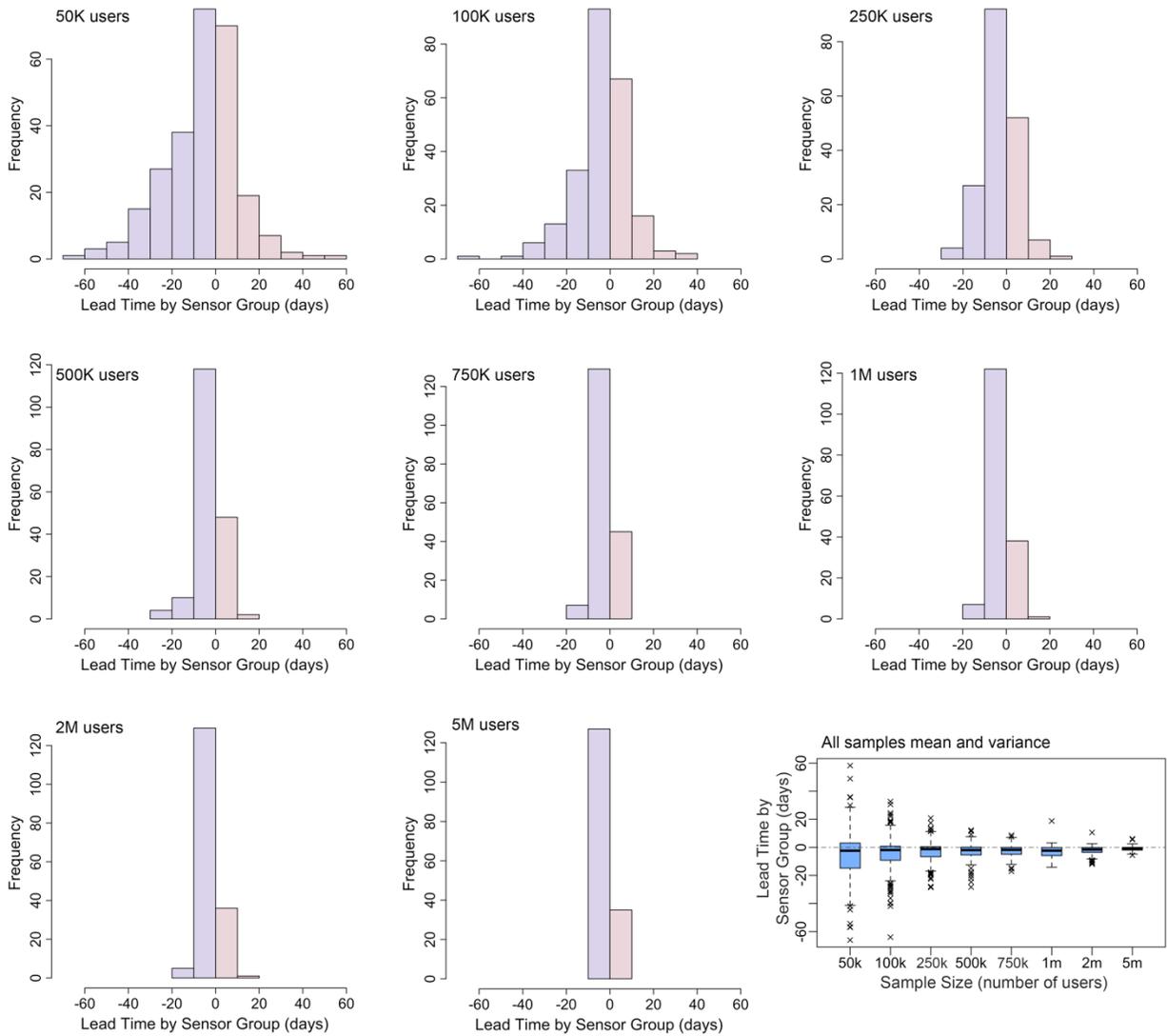

**Figure S5.** Sensor lead time ($\Delta t^{\alpha}$) distribution for hashtags used by more than 0.01% of all users in various sample sizes.

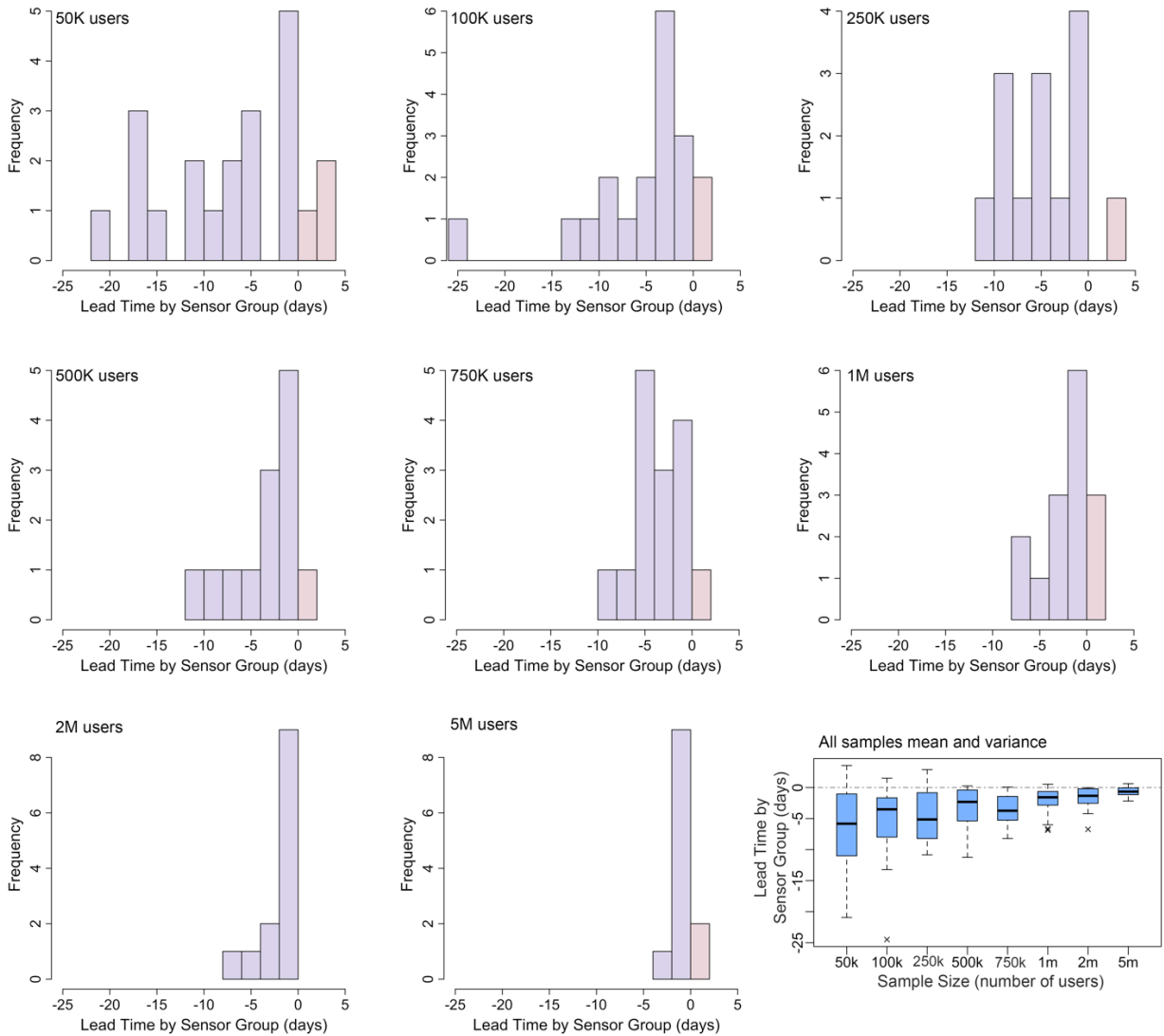

**Figure S6.** Sensor lead time ($\Delta t^\alpha$) distribution for hashtags used by more than 0.04% of users in various sample sizes.

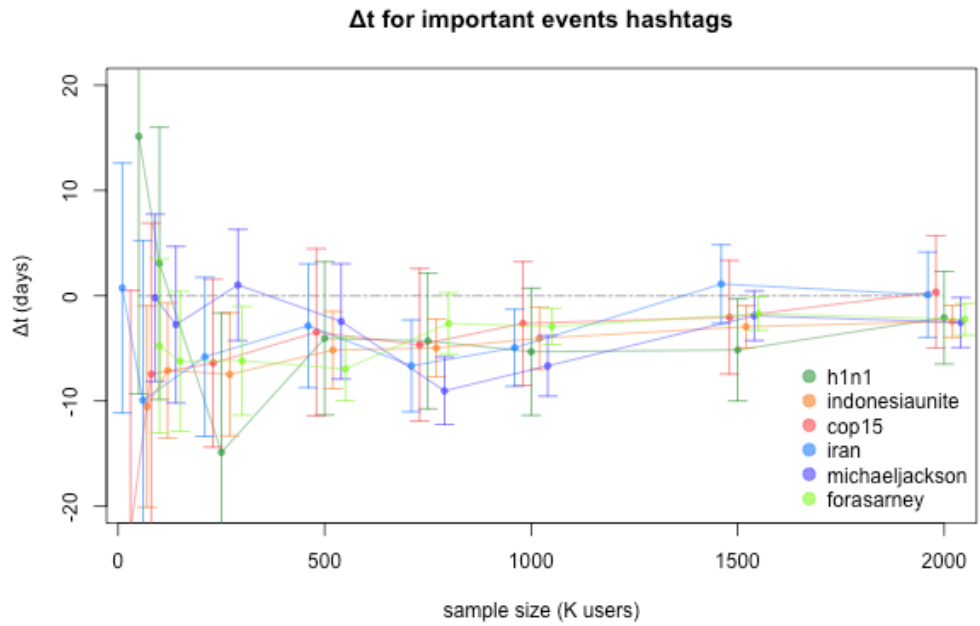

**Figure S7**. Relationship of lead time in the sensor group compared to the control group ($\Delta t^{\alpha}$) in different sample sizes for hashtags related to important events in 2009.

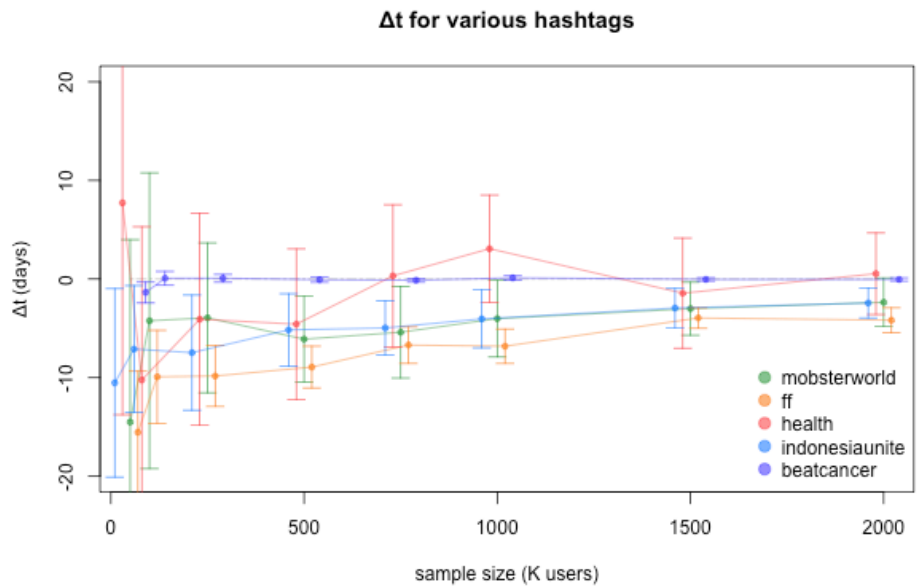

**Figure S8**. Evolution of $\Delta t^\alpha$ with different control samples for *#mobsterworld*, an endogenous viral hashtag; *#ff*, which despite its non virality shows a $\Delta t^\alpha < 0$ similar to that of the simulated infection model on the synthetic network; *#health*, a nonviral hashtag and *#indonesiaunite* and *#beatcancer*, two hashtags for which the sensor mechanism respectively works and does not work.

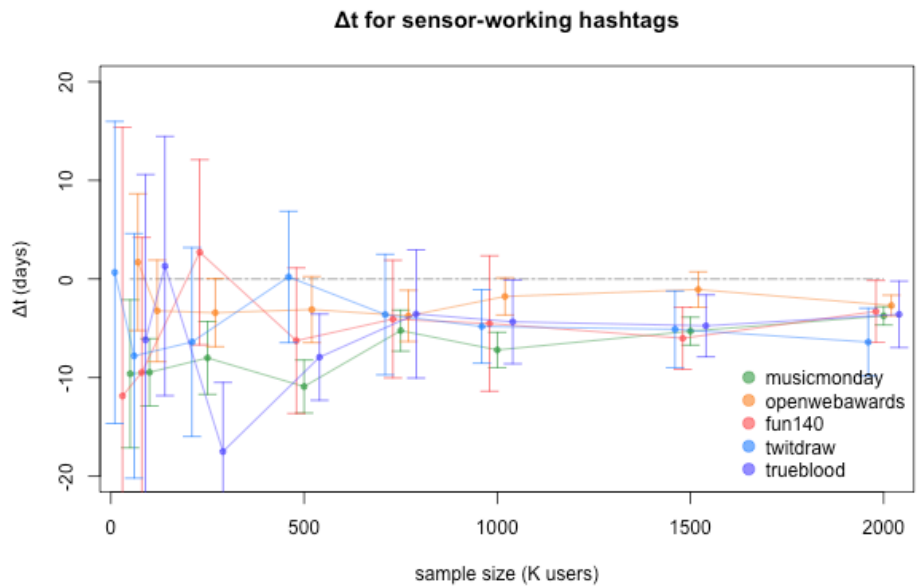

**Figure S9**. Relationship of lead time in the sensor group compared to the control group ($\Delta t^{\alpha}$) in different sample sizes for viral hashtags: *#musicmonday*, *#openwebawards*, *#fun140*, *#twitdraw* and *#forasarney*.

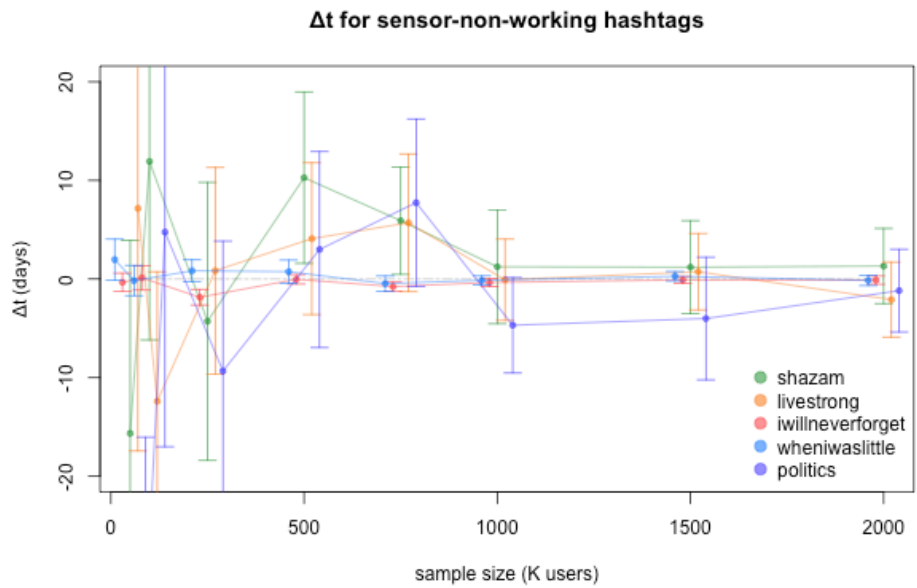

**Figure S10**. Relationship of lead time in the sensor group compared to the control group ($\Delta t^{\alpha}$) in for nonviral hashtags: *#shazam*, *#livestrong*, *#iwillneverforget*, *#wheniwaslittle* and *#politics*.

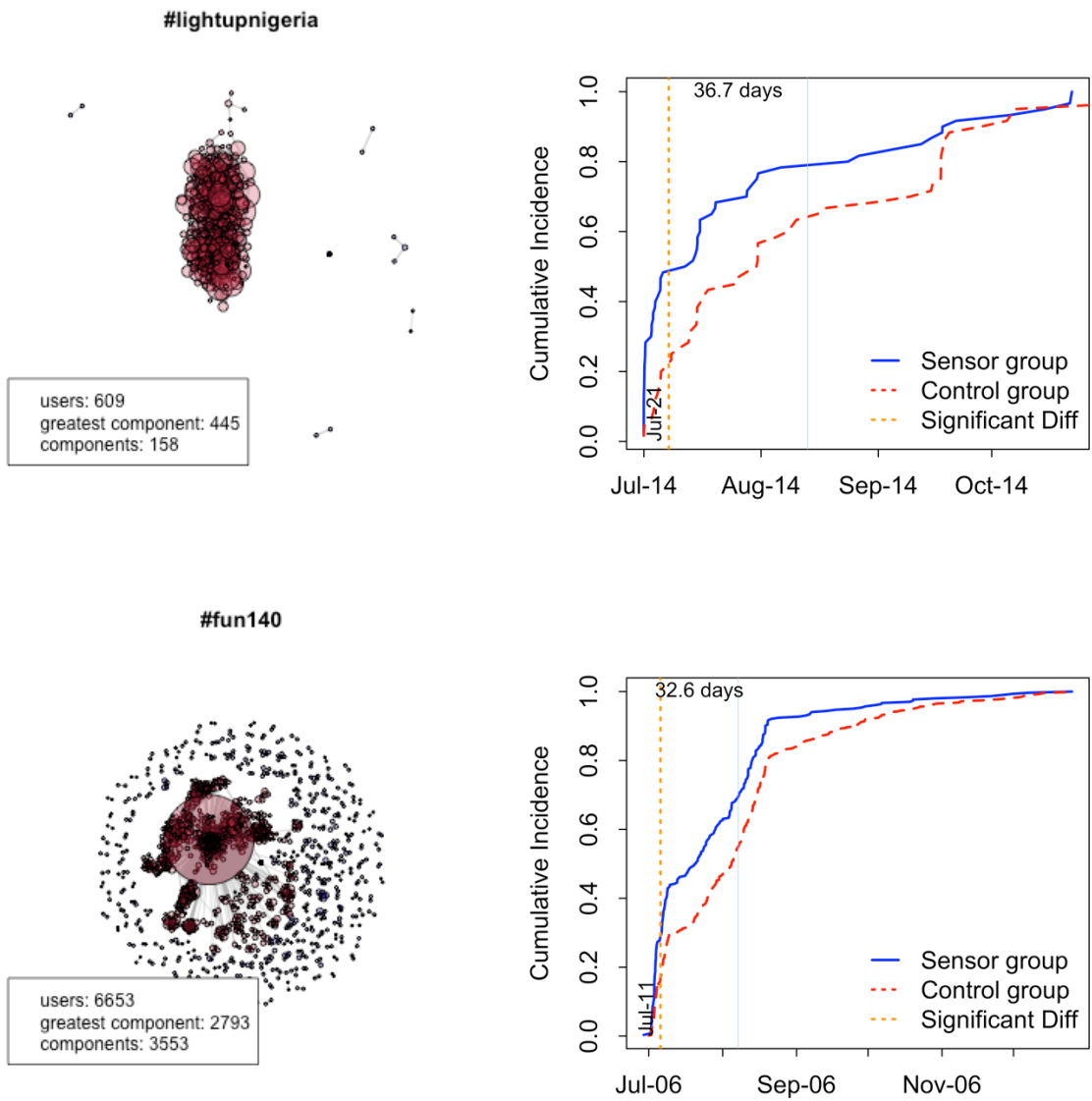

**Figure S11.** Networks and cumulative distributions for the hashtags showing the biggest lead times in Figure 4 of the main text. #lightupnigeria is a small network showing the biggest difference between the cumulative distributions of the control and sensor groups. #fun140 shows the third biggest lead time of the hashtags analyzed and its network displays an unusually large hub at its center.

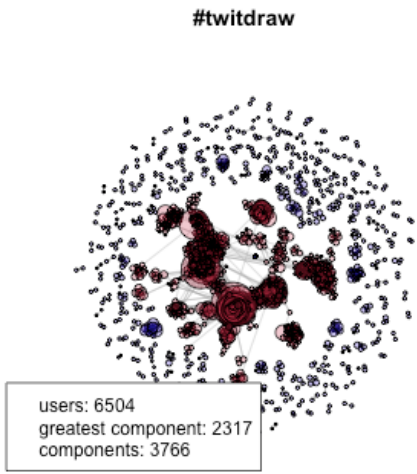
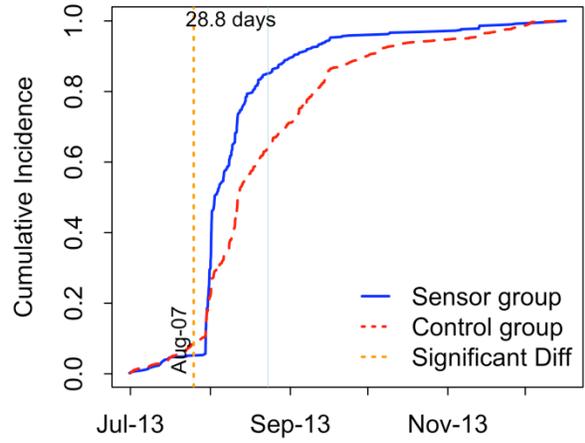

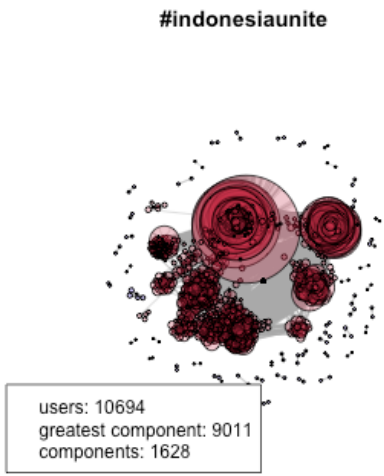
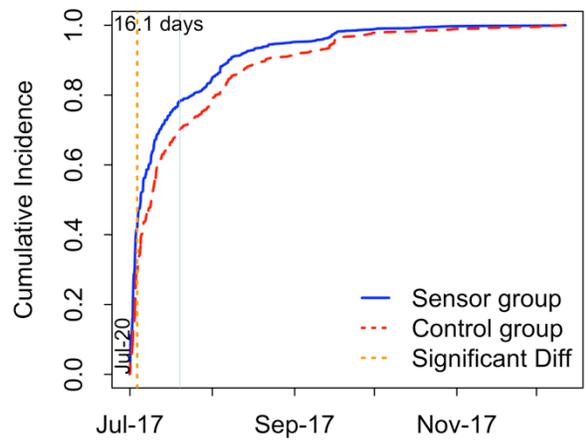

**Figure S12**. Two additional examples showing a considerable lead time for the sensor group.

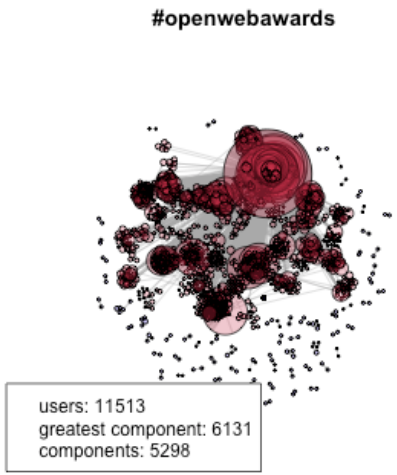
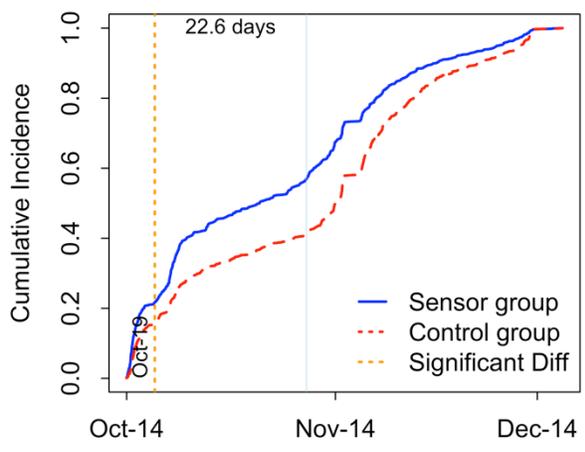

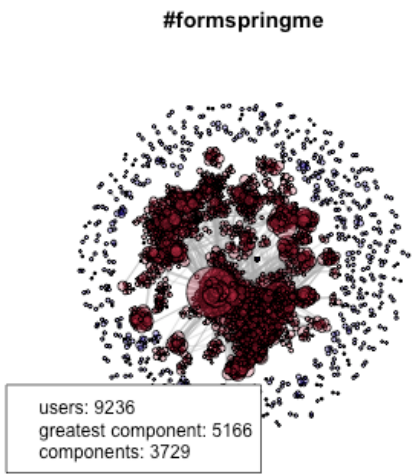
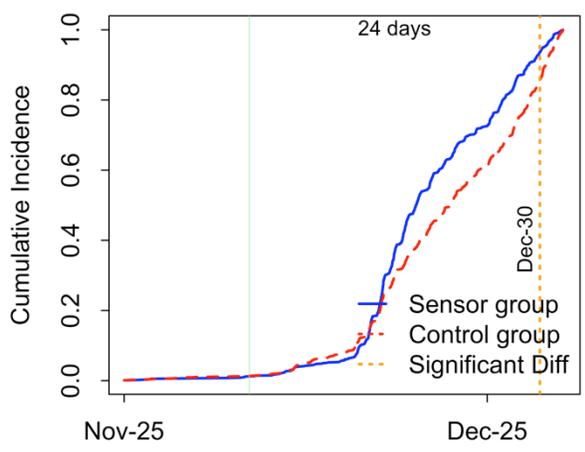

**Figure S13**. Two additional examples showing a considerable lead time for the sensor group.

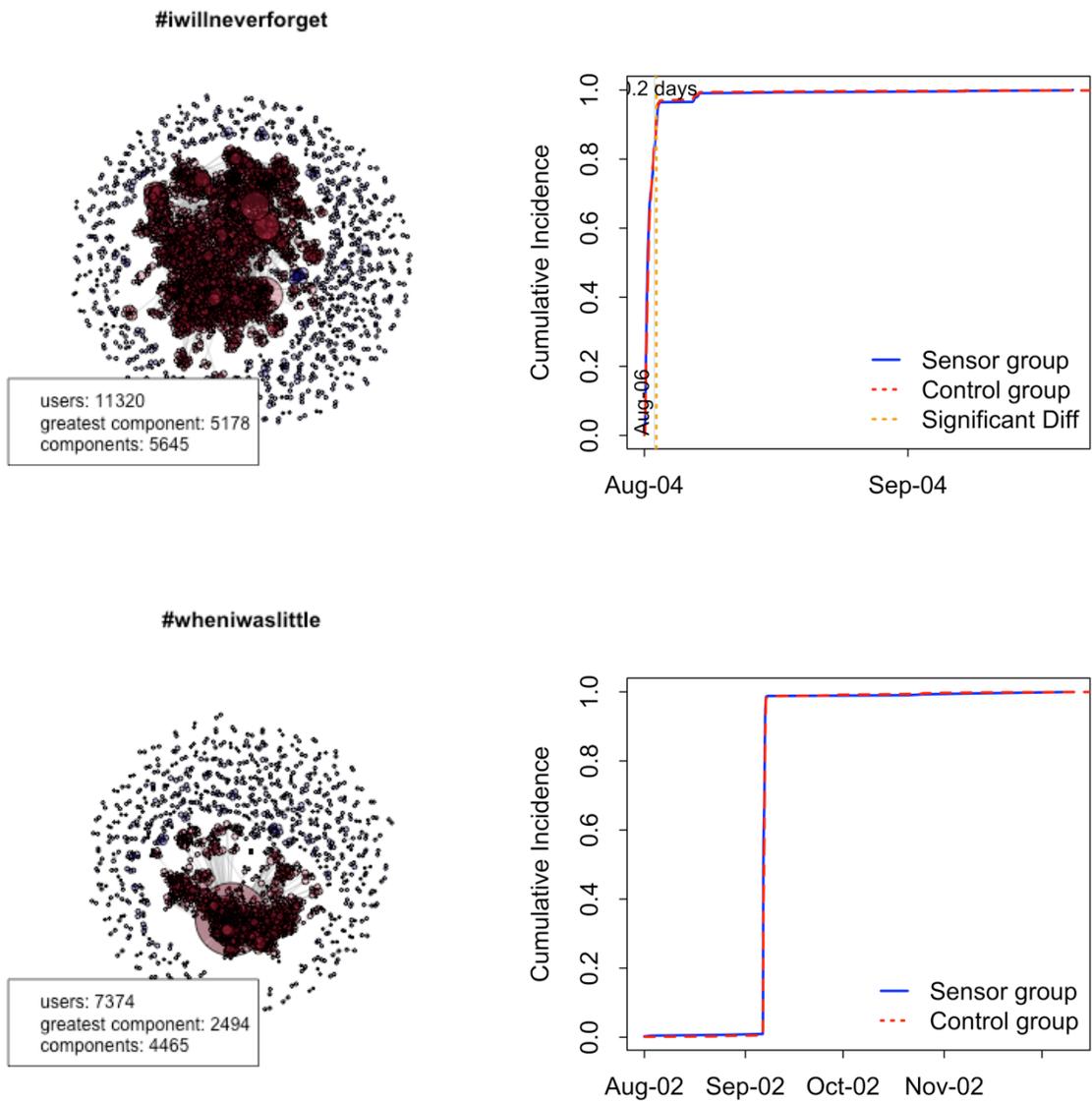

**Figure S14**. Although these two hashtag networks have large giant components, they show signs of very rapid increase in usage, suggesting the sensor mechanism may not give much advance warning in these cases. Their lead times overlap with a those based on randomly shuffling the data as shown in Figure 4 in the main text.

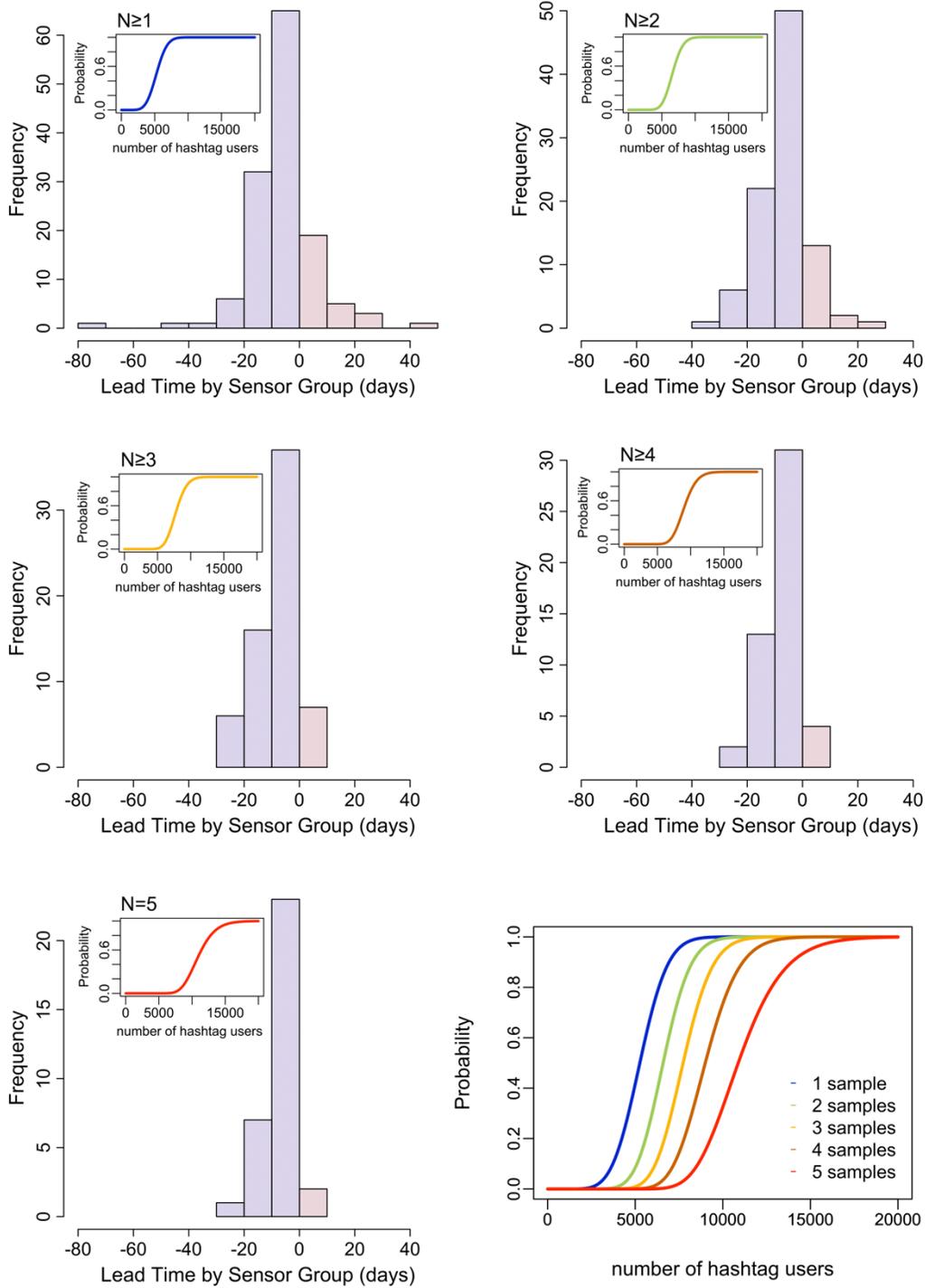

**Figure S15**. Sensor lead time ($\Delta t^{\alpha}$) distribution and total number of users probability distributions for hashtags having at least 10 users in 1 or more of 5 random samples of 50,000 users, 2 or more, 3 or more, 4 or more and in the 5 of them.

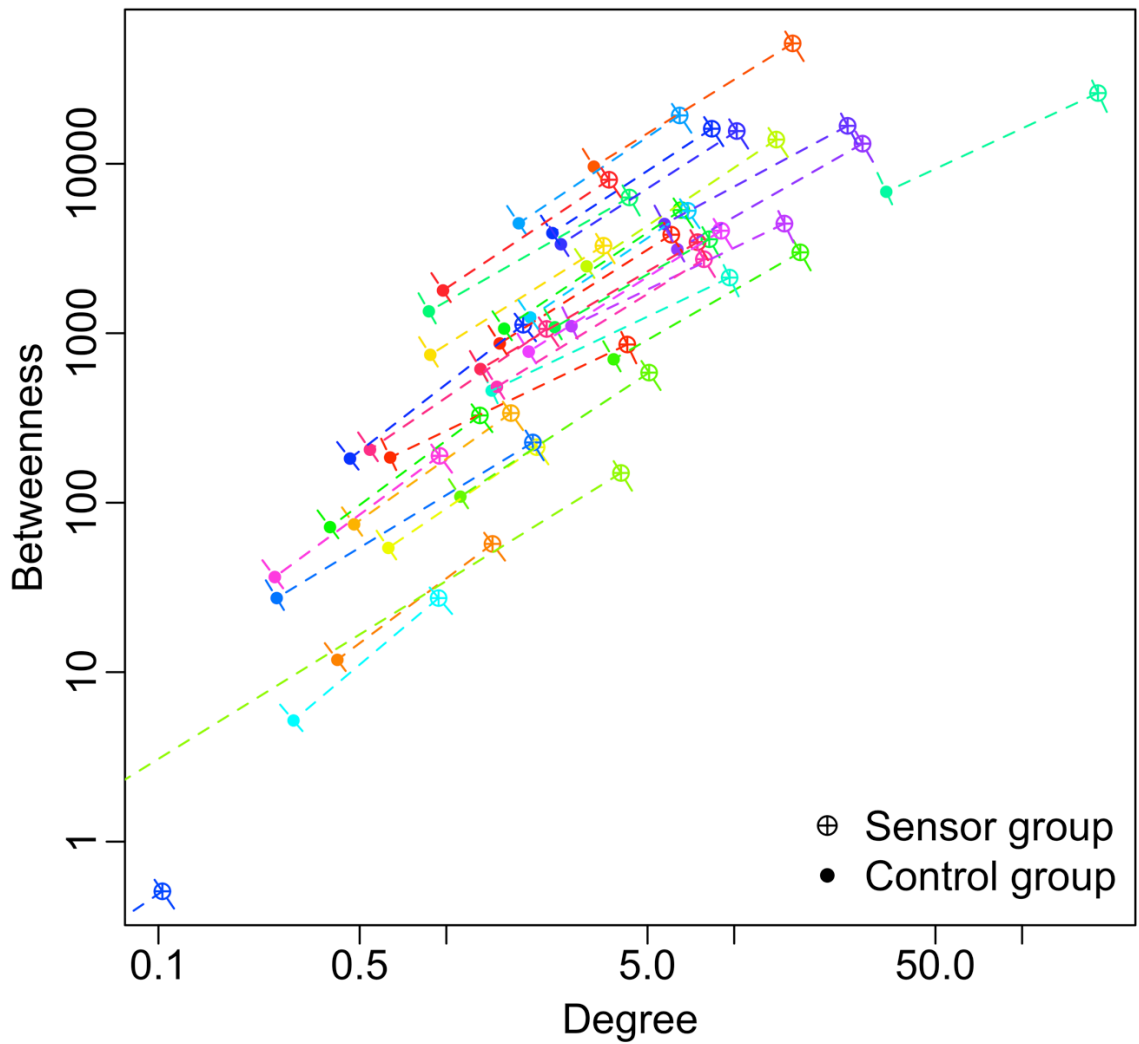

**Figure S16**. Mean degree and betweenness for 340 control groups and their corresponding sensors of 34 different hashtags user networks (those of Figure S6). Each point in the plot correspond to the mean value of 10 different samples of a single hashtag network. Each hashtag network is drawn in a different color. Control and sensor samples are 10% the size of the user network. The sensor mechanism shows to be a reliable method for obtaining not only a sample with higher mean degree but also with higher